\documentclass[twocolumn,english,aps,prd,prl,longbibliography]{revtex4-2}
\usepackage[T1]{fontenc}
\usepackage[a4paper]{geometry}
\usepackage{hyperref}
\usepackage{babel}
\usepackage{amsmath}
\usepackage{tikz}

\begin{document}

\title{Amplitude Bootstrap in (Anti) de Sitter Space And The Four-Point Graviton from Double Copy}

\author{Jiajie Mei$^{a}$}
\affiliation{$^{a}$ Department of Mathematical Sciences, Durham University, Durham, DH1 3LE, UK}

\begin{abstract}
We propose studying a new representation of on-shell Anti de Sitter (AdS) amplitude in Mellin-Momentum space, where it encodes all the dynamical information in Cosmological Correlators. At tree level, we demonstrate that this amplitude has a similar analytic structure as the S-matrix, with residues of poles made up of on-shell lower-point amplitudes. We use this structure to bootstrap 4-point scalar amplitudes with spin-1 and spin-2 exchange. In the second part of the paper, we use double copy to construct the 4-point graviton amplitude in general dimension. This leads us to a novel, concise formula that exhibits the flat space structure. We also verified this formula for the case when d=3 with literature. 
\end{abstract}

\maketitle

\section{Introduction}
Scattering amplitudes are a cornerstone of Quantum Field Theory (QFT),  as they have both theoretical and experimental significance in predicting collider results.  However, defining the S-matrix of QFT in curved spacetime is a challenge. In Anti de Sitter (AdS) space, the gauge-gravity duality allows us to obtain the correlation function of a Conformal Field Theory (CFT) on the boundary\cite{Maldacena:1997re}. In addition, QFT in de Sitter (dS) space offers powerful tools for computing cosmological observables, an active area of research reviewed in \cite{Baumann:2022jpr}.

Over the past decades, there has been tremendous success in bootstrapping the S-matrix using fundamental physical principles such as Lorentz invariance, locality, and unitarity \cite{Benincasa:2007xk,Elvang:2015rqa,Cheung:2017pzi}. For example, the factorization of the four-point massless scalar amplitude with spin-1 and spin-2 exchange completely constrains the amplitude,
\begin{equation}\label{eq:flat1}
A_4 \xrightarrow{P^2 \to 0} \frac{\sum_h A_3^h A_3^{-h}}{P^2},
\end{equation}
where $h$ indicates the possible internal helicity. Generalizing this from flat space to curved space, by replacing Lorentz invariance with conformal invariance, presents considerable challenges. Recent progress has been made in momentum space and cosmology to tackle this problem \cite{Arkani-Hamed:2018kmz,Pajer:2020wxk,Baumann:2020dch,Jazayeri:2021fvk,Goodhew:2020hob}, which is known as the Cosmological Bootstrap. However, cosmological correlators are not invariant under field redefinition/gauge transformation.  While factorization is manifest in momentum space, the pole structure is significantly more intricate, and the special conformal generator in momentum space is a second-order differential operator, making it challenging to implement conformal symmetry. Our proposal in this letter will aim to tackle these challenges in a novel way.

\begin{figure}
  \centering
  \includegraphics[width=1.15\linewidth]{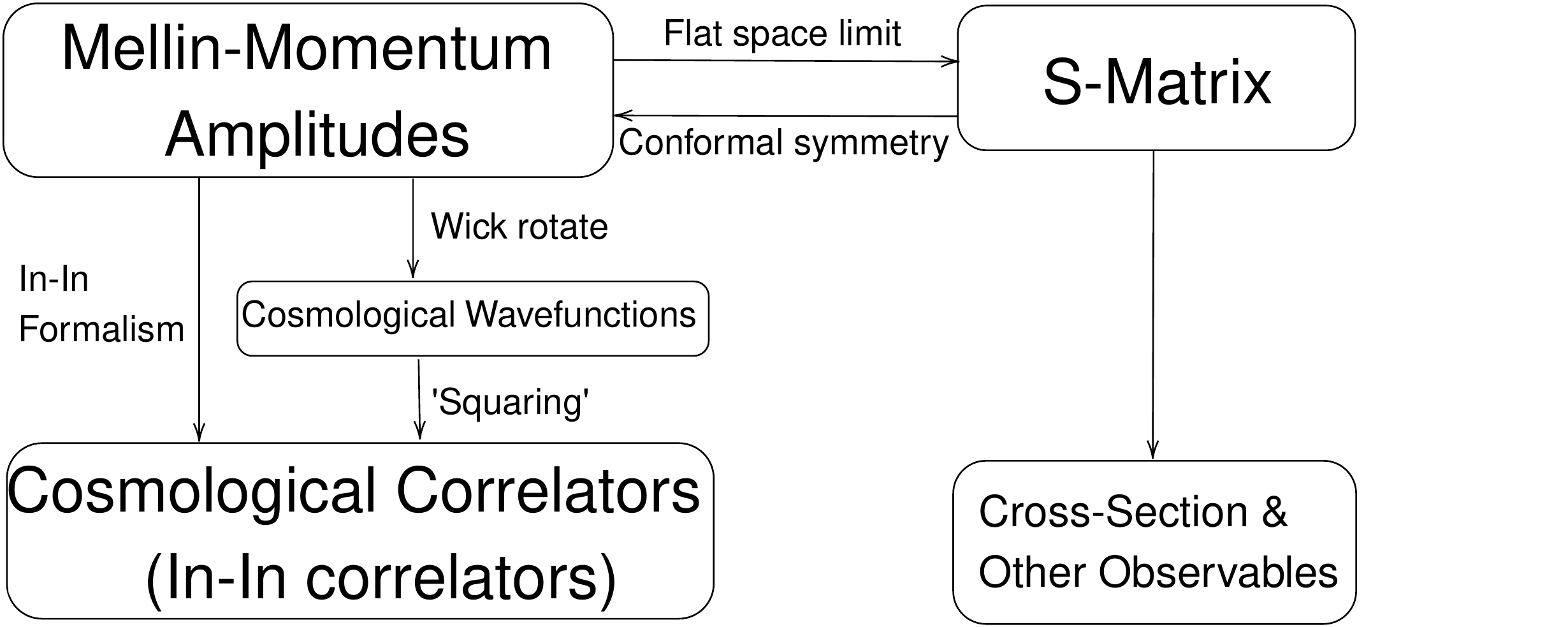}
  \caption{Mellin-Momentum amplitudes in curved space play a similar role to the S-matrix in flat space, with simple analytic structure and the ability to extract observables.}
  \label{fig:relation}
\end{figure}

Inspired by recent developments in differential representation\cite{Eberhardt:2020ewh,Roehrig:2020kck,Gomez:2021qfd,Herderschee:2022ntr,Cheung:2022pdk,Li:2023azu}, perturbiner\cite{Armstrong:2022mfr}, and Mellin-Barnes representation\cite{Sleight:2019hfp,Sleight:2019mgd}, we propose studying a new representation for the AdS amplitude: Mellin-Momentum amplitude. This proposal has two main motivations, described in Fig[\ref{fig:relation}]. Firstly, by  using the recently developed: From AdS to dS \cite{Sleight:2021plv,Sleight:2020obc,DiPietro:2021sjt}, we can easily obtain the observables in Cosmology such as the In-In correlators from AdS amplitudes, as detailed in the Appendix[\ref{Momentum}]. Secondly, at tree level, the residue of the Mellin-Momentum amplitude pole is determined by on-shell lower-point sub-amplitudes, similar to the S-matrix. We now outline how we tackle the challenges in Cosmological Bootstrap \cite{Arkani-Hamed:2018kmz}.
\begin{itemize}
    \item \textbf{On-Shelless}: The amplitude is invariant under field redefinition/gauge transformation \cite{Cheung:2022pdk}.
    \item \textbf{Pole structure}: The amplitude has two types of poles, factorization poles, which are similar to Eq(\ref{eq:flat1}) in flat space, and Operator Product Expansion (OPE) poles, which indicate the exchange of single-particle states from a CFT perspective \cite{Arkani-Hamed:2015bza,Sleight:2019hfp}. 
    \item \textbf{Conformal symmetry}: Instead of solving conformal ward identities, we demand the OPE poles have the correct residue.
\end{itemize}
Demanding factorization limit and OPE limit, we can completely constrain the scalar amplitude with spin-1 and 2 exchange in general dimension.

In the second part of the paper, we will explore the hidden structures in the spinning amplitude for Yang-Mills (YM) and Gravity, such as color/kinematics duality and double copy\cite{Bern:2008qj,Bern:2010ue,Bern:2010yg,Kawai:1985xq,Bern:2019prr}.  As an illustration, we can represent the 4-point Yang-Mills amplitude in flat space schematically as follows,
\begin{equation}
A_4=\frac{n_sc_s}{s}+\frac{n_tc_t}{t}+\frac{n_uc_u}{u},
\end{equation}
where $c_i$ is the standard color factor that satisfies the Jacobi identities $c_s+c_t+c_u=0$, and $n_i$ are known as the kinematic numerators, and obey an analogue Jacobi relation, 
\begin{equation}
n_s+n_t+n_u=0.
\end{equation}
This is known as the kinematic Jacobi identity, which encodes the color/kinematic duality. By replacing the color factor in the Yang-Mills amplitude with the same kinematic numerator, we can obtain the gravity amplitude:
\begin{equation}
M_4=\frac{n_s^2}{s}+\frac{n_t^2}{t}+\frac{n_u^2}{u}.
\end{equation}
This is known as the double copy. We will develop a similar procedure to compute the 4-point graviton amplitude in AdS.

\section{ Mellin-Momentum space}
We work with AdS$_{d+1}$ with radius $\mathcal{R}$  in the Poincar\'e patch, 
\begin{equation}
\tilde{g}_{mn}dx^{m}dx^{n}=\tfrac{\mathcal{R}^{2}}{z^{2}}(dz^{2}+\eta_{\mu\nu}dx^{\mu}dx^{\nu}),
\end{equation}
with $0 <z<\infty$ and $\mathcal{R}=1$. The spacetime indices $m,n,\ldots$ generically represent the radial
direction $z$ and the boundary directions. The latter are denoted
by $\mu,\nu=0,\ldots,d-1$, and $\eta_{\mu\nu}$ is the flat boundary metric. We can make the Wick rotation $z\to-i\eta$ to obtain the analogous computations in dS. Hence the AdS amplitudes consider here can be directly compared with wavefunction coefficients in dS. We will follow the work in \cite{Sleight:2019hfp,Sleight:2019mgd,Sleight:2021plv} to define Mellin-Momentum space. We begin by defining the equation of motion (EoM) operator in momentum space, whose solution without source term gives the scalar bulk-to-boundary propagator:
\begin{align}\label{eq:eom}
    &\mathcal{D}_{k}^{\Delta} \phi_{\Delta} (k,z) =0 ,\nonumber \\
    &\mathcal{D}_{k}^{\Delta}  \equiv   z^{2}k^{2}-z^{2}\partial_{z}^{2}-(1-d)z\partial_{z}+\Delta(\Delta-d),
\end{align}
with  $\Delta$ is scaling dimension and $k=|\vec{k}|$ is norm of boundary momentum. Next, for the spinning particle, we rescale the field accordingly, such that gluon behaves like a $\Delta=d-1$ scalar, while the graviton behaves like a massless scalar\footnote{We restore the $\mathcal{R}$ for dimensional counting.}. To be more specific, for Yang-Mills, we have $\mathbf{A}_{m}=(\mathcal{R}/z) A_{m}$, where $\mathbf{F}_{m n}=\partial_m \mathbf{A}_{n}-\partial_n \mathbf{A}_{m}-i[\mathbf{A}_{m},\mathbf{A}_{n}]$ is the usual field strength. The graviton will be parametrized as $g_{mn}=\tilde{g}_{mn}+\frac{\mathcal{R}^2}{z^2}h_{mn}$, we can then expand this in Einstein field equation and obtain:
\begin{align} 
    \mathcal{D}_{k}^{d-1} A_{\mu}(k,z)&=0 ,\\
    \mathcal{D}_{k}^{d} h_{\mu \nu}(k,z)&=0. 
\end{align}
Clearly, the solutions are just scalar propagators dressed up with boundary polarization.
\begin{align}
    A_{\mu}(k,z)&=\varepsilon_{\mu}\phi_{d-1}(k,z),\\
    h_{\mu \nu}(k,z)&=\varepsilon_{\mu \nu}\phi_{d}(k,z).
\end{align}
Now we consider the Mellin transform of the scalar bulk-to-boundary propagator
\begin{align}
    &\phi_{\Delta}(k,z)=\int_{-i\infty}^{+i\infty} \frac{ds}{2\pi i}z^{-2s+d/2} \phi_{\Delta}(s,k),
\end{align}
where $s$ is the corresponding Mellin variable, similar to boundary momentum: counting the spacetime derivative $(z\partial_z) \phi_{\Delta}=(-2s+d/2)\phi_{\Delta}$. Eq(\ref{eq:eom}) becomes the following,
\begin{align}\label{eq:mellineom}
     (z^{2}k^{2}+(d/2-\Delta)^2-4s^2)\phi_{\Delta}(s,k)=0,
\end{align}
momentum $k$ will always associate with a factor $z$ to capture scale invariance. This is the on-shell condition for the amplitude below. The solution of this equation can be found in: \footnote{To obtain the Mellin-Barnes representation of bulk-to-boundary \cite{Sleight:2021plv}, we can change $z^2k^2 \phi_{\Delta}(s,k)=k^2\phi_{\Delta}(s+1,k)$ to turn it into a difference equation and with appropriate boundary condition, we have:
\begin{equation}
    \phi_{\Delta}(s,k)=\frac{\Gamma(s+\frac{1}{2}(\frac{d}{2}-\Delta))\Gamma(s-\frac{1}{2}(\frac{d}{2}-\Delta))}{2\Gamma(\Delta-\frac{d}{2}+1)}\left( \frac{k}{2} \right) ^{-2s+\Delta-\frac{d}{2}}.
\end{equation}
}.

Now the n-point boundary correlator in momentum space can be written in the following Mellin-Barnes representation:
\begin{align}\label{eq:MMamplitude}
  \langle \mathcal{O}(\Vec{k}_1) \dots \mathcal{O}(\Vec{k}_n)  \rangle & =\delta^d(\Vec{k}_T) \prod_{i=1}^n \int \frac{ds_i}{(2\pi i)}\phi(s_i) A'_n(s), \nonumber \\
  A'_n(s)&=\int \frac{dz}{z^{d+1}} \mathcal{A}_n(z) z^{\sum\limits_{i=1}^{n} (-2s_i+d/2)},
\end{align}
where $\Vec{k}_T=\Vec{k}_1+...+\Vec{k}_n$. We can also integrate out the $z$ variable at every vertex \cite{Sleight:2021plv}:
\begin{align}
  \int \frac{dz}{z^{d+1+b}} z^{\sum\limits_{i=1}^{n} (-2s_i+d/2)}=\delta(d+b+\sum\limits_{i=1}^{n}(2s_i-d/2)),
\end{align}
where $b$ is counting the extra factor of $z$ due to scale invariance, this is referred to Mellin Delta function. Finally, the reduced correlator $\mathcal{A}_n(z)$ is defined as the Mellin-Momentum amplitude, where it encodes all the dynamical information in boundary correlators\footnote{This amplitude can be viewed as the differential representation of the Mellin-Barnes amplitude first proposed in \cite{Sleight:2021iix}. We thank Charlotte Sleight for bringing the references to our attention and sharing the unpublished note.}.\\

\textbf{LSZ reduction formula and local terms}: The boundary correlator in momentum space is not invariant under field redefinition \cite{Maldacena:2011nz,Pajer:2020wxk}. For instance, consider a massless free scalar theory in $d=3$, where under the field redefinition $\phi \to \phi + \alpha \phi^3$, the four-point correlator changes to $\alpha \sum\limits_{i=1}^{4}(k_i^3)$. These are boundary contact terms in momentum space. For Mellin-Momentum amplitude, under such field redefinition, it changes to $\alpha \sum\limits_{i=2}^{4}(z^2  k_1 \cdot k_i +(d/2-2s_1)(d/2-2s_i))$. Importantly, by using boundary momentum conservation, Mellin delta function and on-shell condition Eq(\ref{eq:mellineom}), it vanishes and hence the Mellin-Momentum amplitude is invariant under field redefinition. Similarly, the Conformal Ward identities (CWI) \cite{Bzowski:2013sza} in momentum space, $\sum_{i=1}^{n} D^{AB}_i\Psi_n=...$ \footnote{The $...$ represents the contact terms here, and $AB$ is the notation from Embedding formalism, different component will give different conformal generators.}, only hold up to contact terms. However, the CWI for Mellin-Momentum amplitude is free of contact terms.


These observations can be explained by LSZ reduction formula in AdS with similar argument as QFT in Minkowski space: \emph{contact term is not singular on the on-shell poles}. Contact term of boundary correlator in position space take the following general form:
\begin{align}
    \langle \mathcal{O}(x_1) \dots \mathcal{O}(x_n)  \rangle \delta^{d}(x_i-x_j),
\end{align}
which vanishes unless the two operators collide. We now consider the Mellin-Fourier transform $\phi(x,z)\sim e^{ik\cdot x}z^{-2s+d/2}\phi(s,k)$. Due to the delta function, it becomes a $n-1$ point correlator. Hence, the contact term can not be written as Eq(\ref{eq:MMamplitude}) where the definition needs $\prod_{i=1}^n \phi(s_i,k_i)$.
\begin{quote}
    Therefore, \emph{contact terms in a boundary correlator do not contribute to the Mellin-Momentum amplitude $\mathcal{A}_n$.}
\end{quote}
This is the main reason why we claim that Mellin-Momentum amplitude should be treated as amputated amplitude in AdS.


\section{Factorization limit and OPE limit for scalar four-point amplitude}
In this section, we will adopt the amplitude bootstrap approach to derive the 4-point scalar exchanging spinning particle in AdS. We will demonstrate that due to the simple analytic structure of Mellin-Momentum amplitude, we can replicate the success in the flat space amplitude analysis. We will focus on the spin-1 and spin-2 cases, since they encode the most non-trivial information in YM and gravity amplitude.

At the level of 3-point, by imposing the gauge condition $\varepsilon_i \cdot k_i=0$, dimensional analysis and boundary Lorentz invariance, we can write down the on-shell 3-point amplitude for two scalars and one spin-$\ell$ particle in general dimension:
\begin{align} \label{3ptspinl}
  \mathcal{A}_3=z^{\ell} (k_1 \cdot \varepsilon_3)^{\ell}.
\end{align}
 In general, tree-level Mellin-Momentum amplitudes have two types of poles: factorization poles $\mathcal{D}_{k_s}^{\Delta}$ and OPE poles $\vec{k}_s$, where $\vec{k}_s=\vec{k}_1+\vec{k}_2$. We will begin with spin-1 exchange as an example to demonstrate how studying the pole structure can determine the 4-point amplitude. In the factorization limit:
\begin{align} \label{eq:spin1exchange}
  \mathcal{A}^J_4 \xrightarrow{\mathcal{D}_{k_s}^{d-1} \to 0} & \sum_h \mathcal{A}_3^J \frac{1}{\mathcal{D}_{k_s}^{d-1}}\mathcal{A}_3^J = \frac{z^2\Pi_{1,1}}{\mathcal{D}_{k_s}^{d-1}},
\end{align}
where the sum runs over the possible helicities (guarantee by unitarity) and $J$ denotes conserved current being exchanged. The polarization sums are detailed in Appendix[\ref{polarization sums}]. We stress that the inverse operator should act on the three-point amplitude and the inverse is explicitly defined in Appendx[\ref{Momentum}].

We now turn to the OPE limit, taking $\Vec{k}_s \to 0$. This limit is determined by the Conformal Partial Wave (CPW)\cite{Sleight:2019hfp}\footnote{where $\hat{\partial_{\varepsilon}}$ is the Todorov operator on the boundary, and $\tilde{\mathcal{O}}$ is the shadow operator.}
\begin{align} \label{CPW}
 & \lim_{\vec{k}_{s} \to 0}  \langle \mathcal{O}(\Vec{k}_1)\mathcal{O}(\Vec{k}_2)\mathcal{O}(\Vec{k}_3)\mathcal{O}(\Vec{k}_4) \rangle \nonumber \\
  \sim &  \langle \mathcal{O}(\Vec{k}_1) \mathcal{O}(\Vec{k}_2)  \mathcal{O}(\Vec{k}_s;\hat{\partial_{\varepsilon}}) \rangle \langle \tilde{\mathcal{O}}(-\Vec{k}_s;\varepsilon) \mathcal{O}(\Vec{k}_3)\mathcal{O}(\Vec{k}_4) \rangle.
\end{align}
Putting the 3-point data Eq(\ref{3ptspinl}) into the equation above implies that,
\begin{align} \label{eq:spin1ope}
  \mathcal{A}^J_4 \xrightarrow{\vec{k}_s \to 0} & \mathcal{O}(k_s^0).
\end{align}
This is all we need from CPW.
The leading term in OPE limit from Eq(\ref{eq:spin1exchange}) is controlled by,
\begin{align}
  \frac{z^2(k_1^2-k_2^2)(k_3^2-k_4^2)}{16k_s^2 (s_{12}-d/2)(s_{12}-1)},
\end{align}
where $s_{12}=s_{1}+s_{2}$. In the denominator, we replaced the $z$ derivative with Mellin variables because in the OPE limit it behaves like contact interaction. Then by using EoM Eq(\ref{eq:mellineom}) and Mellin delta function, it becomes:
\begin{align}
  -\frac{(s_1-s_2)(s_3-s_4)}{z^2k_s^2 }.
\end{align}
 Amazingly, the Mellin variables in the denominator cancel exactly. On the other hand, Eq(\ref{eq:spin1ope}) tells us that pole $(k_s)^{-2}$ must cancel exactly. Hence, this term should then be subtracted (to have the correct OPE limit), therefore completely fixed the 4-point amplitude:
\begin{align}
  \mathcal{A}^J_4 = \frac{z^2\Pi_{1,1}}{\mathcal{D}_{k_s}^{d-1}} +\Pi_{1,0}.
\end{align}
Next, we will turn to minimally coupled scalars exchanging graviton. The factorization limit demands that
\begin{align}
  \mathcal{A}^T_4 \xrightarrow{\mathcal{D}_{k_s}^{d} \to 0} \sum_h \mathcal{A}_3^T \frac{1}{\mathcal{D}_{k_s}^{d}}\mathcal{A}_3^T= \frac{z^4\Pi_{2,2}}{\mathcal{D}_{k_s}^{d}},
\end{align}
where $T$ denotes the stress tensor being exchanged. In the OPE limit, the situation becomes slightly more complicated because of the additional terms. Details can be found in Appendix[\ref{polarization sums}] along with the definition of $\Pi_{2,i}$.\\
In the end, we fixed the four-point scalar with graviton exchange as follows:
\begin{align}
  \mathcal{A}^T_4= \frac{z^4\Pi_{2,2}}{\mathcal{D}^d_{k_s}} +z^2\Pi_{2,1} +\Pi_{2,0}.
\end{align}
We have also verified this formula with literature, details can be found in Appendix[\ref{Momentum}].

\section{Yang-Mills Amplitude and Color/Kinematics Duality}
In the following sections, we will tackle the most interesting cases of AdS amplitudes: Yang-Mills and Gravity. While spinor-helicity techniques have shown to be very efficient in the flat space amplitude bootstrap, it is still unclear how to achieve similar simplicity in the AdS context. So in the rest of this letter, our strategy will be writing the known expression for Yang-Mills from \cite{Armstrong:2022mfr} into Mellin-Momentum space, and exploiting double copy to compute the gravity amplitude.

The 3-point color-ordered YM amplitude in general dimension takes the following form (This Mellin form and 3-point Gravity double copy below was first presented in \cite{sleight2020exploring})
\begin{align}
    \mathcal{A}_3=z(\varepsilon_1 \cdot \varepsilon_2 \varepsilon_3 \cdot k_1+\varepsilon_2 \cdot \varepsilon_3 \varepsilon_1 \cdot k_2 + \varepsilon_3 \cdot \varepsilon_1 \varepsilon_2 \cdot k_3).
\end{align}
Next, for the 4-point Yang-Mills amplitude we have
\begin{align}
    \mathcal{A}_4=&\frac{z^2\varepsilon_1 \cdot \varepsilon_2 \varepsilon_3 \cdot \varepsilon_4  \Pi_{1,1}+z^2 W_s }{\mathcal{D}_{k_s}^{d-1}} +\varepsilon_1 \cdot \varepsilon_2 \varepsilon_3 \cdot \varepsilon_4 \Pi_{1,0} +V_c^s \nonumber \\
    &-[(12)\to(23)],\label{eq:4pt-gluon}
\end{align}
where
\begin{align}
    4W_s=&\varepsilon_1 \cdot \varepsilon_2 (k_1 \cdot \varepsilon_3 k_2 \cdot \varepsilon_4 -k_2 \cdot \varepsilon_3 k_1 \cdot \varepsilon_4 )   \nonumber \\
    +&\varepsilon_3 \cdot \varepsilon_4 (k_3 \cdot \varepsilon_1 k_4 \cdot \varepsilon_2 -k_4 \cdot \varepsilon_1 k_3 \cdot \varepsilon_2 )   \nonumber \\
    +&(k_2 \cdot \varepsilon_1 \varepsilon_2 -k_1 \cdot \varepsilon_2  \varepsilon_1 ) \cdot (k_4 \cdot \varepsilon_3 \varepsilon_4 -k_3 \cdot \varepsilon_4  \varepsilon_3 ) ,
\end{align}
and the s-channel contact diagram is
\begin{align}
    4V_c^s=&\varepsilon_1 \cdot \varepsilon_3 \varepsilon_2 \cdot \varepsilon_4 -\varepsilon_1 \cdot \varepsilon_4 \varepsilon_2 \cdot \varepsilon_3 .
\end{align}
We can now extract the kinematic numerator from the expression above:
\begin{align}
    n_s=&\varepsilon_1 \cdot \varepsilon_2 \varepsilon_3 \cdot \varepsilon_4 (z^2\Pi_{1,1}+\Pi_{1,0}\mathcal{D}_{k_s}^d) +z^2W_s+ V_c^s \mathcal{D}_{k_s}^d, \label{eq:kinematic numerator}
\end{align}
where we have replaced the EoM operator Eq(\ref{eq:eom}) with the different conformal dimension for Gravity. This seems to be an unavoidable procedure in curved space and we will discuss more about it in section\ref{sec:final}. 
The reason we keep the kinematic numerator as an operator form is so that when we perform the double copy, we can simply cancel it with the propagator in the denominator. However, the EoM operator itself by definition Eq(\ref{eq:MMamplitude}) is equivalent to $\mathcal{D}_{k_s}^d=z^2k_s^2+4s_{12}s_{34}$. It's noteworthy that after this replacement the kinematic numerator is free of the OPE pole $k_s$ now. Other channels can be obtained by permutation:
\begin{equation}
    n_t=n_s\bigl|_{(234)\to(423)}, \quad  n_u=n_s\bigl|_{(234)\to(342)}.
\end{equation}
It's easy to verify that the kinematic Jacobi identity is satisfied, similar to the case of flat space amplitude\cite{Armstrong:2020woi,Albayrak:2020fyp,Alday:2021odx,Diwakar:2021juk,Li:2022tby,Drummond:2022dxd}:
\begin{equation}\label{eq:C/K}
    n_s+n_t+n_u =0.
\end{equation}

\section{Graviton Amplitude And Double copy}
Finally, we are ready to discuss the most important example in this letter, namely the gravity amplitude. Firstly, the 3-point gravity amplitude:
\begin{align}
    \mathcal{M}_3=z^2(\varepsilon_1 \cdot \varepsilon_2 \varepsilon_3 \cdot k_1+\varepsilon_2 \cdot \varepsilon_3 \varepsilon_1 \cdot k_2 + \varepsilon_3 \cdot \varepsilon_1 \varepsilon_2 \cdot k_3)^2.
\end{align}
This has a manifestly double copy structure \cite{Farrow:2018yni,Lipstein:2019mpu,Li:2022tby,Lee:2022fgr,sleight2020exploring,Li:2018wkt,Caron-Huot:2021kjy} with the appropriate normalization:
\begin{align}
    \mathcal{M}_3=(\mathcal{A}_3)^2.
\end{align}
This relation has no ordering ambiguity and is valid in general dimensions. However, this is not the full story. As double copy of pure Yang-Mills should give graviton coupled to dilaton and antisymmetric tensor. Considering the tensor product of the polarization, we decompose it into a transverse and traceless tensor (graviton) and a trace (dilaton),
\begin{align}
    \varepsilon^{\mu}\varepsilon^{\nu}=\frac{1}{2}(\varepsilon^{\mu}\varepsilon^{\nu}+\varepsilon^{\nu}\varepsilon^{\mu}-\frac{2}{d-1}\Pi^{\mu \nu})+\frac{1}{d-1}\Pi^{\mu \nu},
\end{align}
where $\Pi_{\mu \nu}=\eta_{\mu \nu}-\frac{k_{\mu}k_{\nu}}{k^2}$ is the projection tensor. This predicts a new interaction between two graviton $h_{\mu \nu}$ and one dilaton $\phi$ (where dilaton is identified as $\varepsilon^+ \varepsilon^-$):
\begin{align}
    \mathcal{M}_3(1_h,2_h,3_{\phi})&=(\mathcal{A}_3)^2|_{\varepsilon_3^{\mu}\varepsilon_3^{\nu}\to \Pi^{\mu \nu}}, \nonumber \\
    &=z^2(\varepsilon_1 \cdot \varepsilon_2)^2k_1^{\mu}k_1^{\nu}\Pi_{\mu \nu}.
\end{align}
This amplitude has a vanishing flat space limit as expected. Moving to four-point, with the color/kinematic duality satisfying in Eq(\ref{eq:C/K}), we can replace the color factor with the kinematic numerator\cite{Armstrong:2023phb,Lipstein:2023pih,Zhou:2021gnu,Bissi:2022wuh,Li:2022tby},
\begin{align}
    \mathcal{M}_4 = \frac{n_s^2}{\mathcal{D}_{k_s}^d}+\frac{n_t^2}{\mathcal{D}_{k_t}^d}+\frac{n_u^2}{\mathcal{D}_{k_u}^d}.\label{eq:4pt-gravitondcg}
\end{align}
However, with the new 3-point interaction found above, the double copy result will include four external graviton exchanging dilaton. It would be very interesting to understand whether Eq(\ref{eq:4pt-gravitondcg}) corresponds to the four-graviton amplitude in a dilaton-graviton theory. We leave this exploration to the future. Instead, here we will extract Einstein gravity from the double copy above by using a similar strategy in flat space, for examples, pure gravity at loop level and massive scalar\cite{Carrasco:2021bmu,Kosower:2022yvp,Johansson:2014zca,Luna:2017dtq,Carrasco:2020ywq,Plefka:2019wyg}. To project out the dilaton scalar degree of freedom, we can demand the factorization only has graviton propagation:
\begin{align}
    \mathcal{M}_4^{EG} = \frac{n_s^2}{\mathcal{D}_{k_s}^d}+\frac{n_t^2}{\mathcal{D}_{k_t}^d}+\frac{n_u^2}{\mathcal{D}_{k_u}^d}-\tilde{\mathcal{M}}_{\mathrm{AdS}}.\label{eq:4pt-gravitondc}
\end{align}
So we subtracted out the dilaton state,
\begin{align}\label{eq:4ptgravitycorrection}
    \tilde{\mathcal{M}}_{\mathrm{AdS}}=&(\varepsilon_1 \cdot \varepsilon_2 \varepsilon_3 \cdot \varepsilon_4 )^2(\frac{z^4\Pi_{2,2}^{\mathrm{Tr}}}{\mathcal{D}_{k_s}^d}-\Pi_{2,0}+\Pi_{1,0}^2\mathcal{D}_{k_s}^d) \nonumber \\
    &+\mathcal{P}(2,3,4),
\end{align}
where $\mathcal{P}(2,3,4)$ denotes sum over permutation to obtain the t-channel and u-channel. The first correction term is the dilaton exchange and the last two terms can be understood as the conformal structure of the graviton propagator. This formula is the first 4-point gravity amplitude in $\mathrm{AdS}_{d+1}$ and takes on a remarkably simple form. In particular, this formula exhibits flat space structure and explains the origin of the complex contact interaction terms as simply zero/two derivative scalar contact interaction like flat space amplitude. As a result, one can simply replace the flat space amplitude by Eq(\ref{eq:upliftingop}) to obtain AdS amplitude. We have explicitly verified that it matches with \cite{Bonifacio:2022vwa} in $d=3$ by reverting back to momentum space.\\
\section{Final remarks} \label{sec:final}

\textbf{Remark on flat space structure}: Given the simplicity of the Mellin-Momentum amplitude and its resemblance to its flat space counterpart, we can define an uplifting operation as follows:
\begin{align} \label{eq:upliftingop}
    \mathcal{U}: \{ & \epsilon_i \cdot \epsilon_j \to \varepsilon_i \cdot \varepsilon_j ,\epsilon_i \cdot k_j \to z \varepsilon_i \cdot k_j, \nonumber \\
    &\frac{(T-U)^2}{S}  \to \frac{z^4\Pi_{2,2}}{\mathcal{D}^{\Delta}_{k_s}}+\Pi_{2,1}+\Pi_{2,0},\nonumber \\
    &\frac{T-U}{S} \to \frac{z^2\Pi_{1,1}}{\mathcal{D}^{\Delta}_{k_s}}+\Pi_{1,0}, S \to z^2k_s^2+4s_{12}s_{34} \} ,
\end{align}
which uplifts all the scattering variables in $(d+1)$ Minkowski space to the AdS ones. In hindsight, the last three steps are essentially stating that we should replace Lorentz-invariant quantities with conformally-invariant ones. It is noteworthy that all the examples considered in this letter adhere to this form,
\begin{align}
    \mathcal{A}^{\mathrm{AdS}}=\mathcal{U} (A^{\mathrm{Mink}}).
\end{align}
It would be interesting to compare this operation with weight-shifting operators approach:\cite{Baumann:2019oyu,Li:2022tby,Lee:2022fgr,Bonifacio:2021azc,Baumann:2021fxj}.\\
\textbf{Remark on flat space limit}: Based on the previous work \cite{Penedones:2010ue,Raju:2012zr}, it is easy to guess that the flat space limit of the Mellin-Momentum amplitude can be obtained by taking the scaling limit of the Mellin variables $s_i \to \infty$, and then replacing them with the corresponding norm of momentum:
\begin{align}
    \lim_{\mathcal{R}_{\rm AdS}\rightarrow\infty,s_i \to \infty} \mathcal{A}^{\mathrm{AdS}} \xrightarrow{s_i \to \frac{zk_i}{2}} A^{\mathrm{Mink}} \delta^{d+1}(\sum_{i}^n \Vec{k}_i).
\end{align}
Under the scaling limit, the Mellin mode in $z$ direction behaves like a Fourier mode and hence the delta function in flat space naturally arises from combining the boundary momentum conservation with the Mellin delta function. One might try to prove this formula following the discussion in \cite{Li:2021snj}.\\
\textbf{Remark on Double Copy in curved space}: In our proposal, the squaring process naturally mimics the Double Copy structure in the S-matrix, while the rest of the bootstrap procedure aims to probe the extra structure in curved space. The additional constraint we need to construct the Double Copy to Gravity might be a generic feature in curved spacetime. The color/kinematic duality and double copy for Non-Linear Sigma Models(NLSM) was studied in \cite{Sivaramakrishnan:2021srm,Diwakar:2021juk,Cheung:2022pdk,Armstrong:2022csc,Armstrong:2022vgl}. Moreover, in \cite{Cheung:2022pdk} the authors showed that the duality holds off-shell at symmetric spacetime manifold. However, the process of replacing color with kinematic is blind to the extra quantum number of conformal dimension, which has different values in AdS for NLSM, and sG \cite{Armstrong:2022vgl,Bonifacio:2018zex}.

\textbf{Future directions}:
 An important direction would be to consider higher-point amplitudes. Indeed, in \cite{Mei:2024abu} we have found the 5-point tree-level YM amplitude has a similar flat space structure. With the on-shell constructibility and flat space structure shown here, another natural question to ask is that whether we can construct on-shell recursion \cite{Britto:2005fq} in AdS. It would be very interesting to better understand the new interaction predicted by double copy, especially it may shed new light on double copy/KLT relation \cite{Kawai:1985xq} to string amplitudes in curved space. The recent discovery of single-valued zeta functions in the leading curvature correction to the AdS Virasoro-Shapiro amplitude \cite{Alday:2023jdk,Alday:2022xwz} strongly indicates the possibility of such relation.

\begin{acknowledgments}
We would like to thank  Luis F. Alday, Connor Armstrong, Harry Goodhew, Renann Lipinski Jusinskas, Guanda Lin, Yue-Zhou Li, Paul McFadden, Silvia Nagy, Enrico Pajer for valuable discussions and comments on the draft, and especially Arthur Lipstein and Charlotte Sleight for many valuable comments and suggestions to improve the draft during the completion of this work. JM is supported by a Durham-CSC Scholarship.
\end{acknowledgments}

\bibliography{refs.bib} 

\pagebreak
\begin{widetext}
\appendix

\section{Polarization sums } \label{polarization sums}
In this appendix, we provide the details of the polarization sums employed in this letter. Following the boundary transverse gauge \cite{Armstrong:2022mfr}: (This is the same as in QFT textbook \cite{Weinberg:1995mt} with Coulomb gauge.)
\begin{align}
    \sum_{h=\pm} \varepsilon_{\mu}(k,h) \varepsilon_{\nu}(k,h)^* &=\eta_{\mu \nu}-\frac{k_{\mu}k_{\nu}}{k^2} \equiv \Pi_{\mu \nu}, \\
    \sum_{h=\pm} \varepsilon_{\mu \nu}(k,h) \varepsilon_{\rho \sigma}(k,h)^* &=\frac{1}{2} \Pi_{\mu \rho}\Pi_{\nu \sigma}+ \frac{1}{2}\Pi_{\mu \sigma} \Pi_{\rho \nu}-\frac{1}{d-1}  \Pi_{\mu \nu}\Pi_{\rho \sigma},
\end{align}
which are transverse and traceless projection tensor. Let's return to QED in Coulomb gauge for a moment. The polarization tensor above which appear in the photon propagator is not Lorentz invariant on its own, but we can restore Lorentz invariance to obtain the covariant photon propagator. This is the same logic that we use to derive all of the polarization sums below by demanding conformal invariance.\\
Finally, let us explicitly write out the polarization sums at 4-point, see also \cite{Arkani-Hamed:2018kmz,Baumann:2020dch,Baumann:2021fxj} for the case of conformally
coupled scalar.
\begin{align}
\Pi_{1,1}\equiv &\frac{1}{4} (k_{1}^{\mu}-k_2^{\mu})\Pi_{\mu \nu}(k_{3}^{\nu}-k_4^{\nu})=\frac{1}{4}(k_1-k_2)\cdot (k_3-k_4)+\frac{(k_1^2-k_2^2)(k_3^2-k_4^2)}{4k_s^2}, \\
\Pi_{1,0}\equiv & -\frac{(s_1-s_2)(s_3-s_4)}{z^2 k_s^2}.
\end{align}
Next, we write the spin-2 polarization sums in a way that makes its double copy structure clear.
\begin{align}
\Pi_{2,2}\equiv & \frac{1}{16}(k_{1}^{\mu}-k_2^{\mu})(k_{1}^{\nu}-k_2^{\nu})(\frac{1}{2} \Pi_{\mu \rho}\Pi_{\nu \sigma}+ \frac{1}{2}\Pi_{\mu \sigma} \Pi_{\rho \nu}-\frac{1}{d-1}  \Pi_{\mu \nu}\Pi_{\rho \sigma})(k_{3}^{\rho}-k_4^{\rho})(k_{3}^{\sigma}-k_4^{\sigma})\\
=&\Pi_{1,1}^2-\Pi_{2,2}^{\mathrm{Tr}}, \\
\Pi_{2,1}\equiv & 2\Pi_{1,1} \Pi_{1,0},
\end{align}
\begin{align}
\Pi_{2,2}^{\mathrm{Tr}}\equiv & \frac{(k_{1}^{\mu}-k_2^{\mu})\Pi_{\mu \nu}(k_{1}^{\nu}-k_2^{\nu})(k_{3}^{\rho}-k_4^{\rho})\Pi_{\rho \sigma}(k_{3}^{\sigma}-k_4^{\sigma})}{16(d-1)}, \\ 
\Pi_{2,0}\equiv & -\frac{d(k_1^2-k_2^2)(k_3^2-k_4^2)(s_1-s_2)(s_3-s_4)}{4(d-1)k_s^4} +\frac{(d-2)z^2(k_1^2-k_2^2)(k_3^2-k_4^2)(s_1-s_2)(s_3-s_4)}{4(d-1)k_s^2(d-2s_{12})(d-2s_{34})} \nonumber \\
&+\frac{(k_1^2-k_2^2)(s_1-s_2)(d^2-8(s_3^2+s_4^2))}{8(d-1)k_s^2(d-2s_{12})}+\frac{(k_3^2-k_4^2)(s_3-s_4)(d^2-8(s_1^2+s_2^2))}{8(d-1)k_s^2(d-2s_{34})} \nonumber \\
&+\frac{(z^2k_s^2+4s_{12}s_{34})}{(d-1)}+\frac{4(s_1-s_2)^2+4(s_3-s_4)^2-d^2}{16(d-1)}.
\end{align}

Both $\Pi_{2,1}$ and $\Pi_{2,0}$ can be determined in the same way as we obtained $\Pi_{1,0}$ in the main text. Note that there are still terms with Mellin variables in the denominator which naively violate locality. However, they will all cancel after using the Mellin delta function.\\

\section{Back to Momentum space and Cosmological Correlators\label{Momentum}}

In this letter, we focused on the analytic structure of Mellin-Momentum amplitude. However, it is also important to stress that we can easily obtain the actual observables: Cosmological correlators. As a non-trivial example, we will give a detailed translation from the Four-point gravity amplitude to the gravity Trispectrum \cite{Bonifacio:2022vwa}. Expanding out the full expression from Eq(\ref{eq:4pt-gravitondc}):
\begin{align}
 &\mathcal{M}_4=\frac{(\varepsilon_1 \cdot \varepsilon_2 \varepsilon_3 \cdot \varepsilon_4 z^2 \Pi_{1,1}+z^2 W_s)^2-(\varepsilon_1 \cdot \varepsilon_2 \varepsilon_3 \cdot \varepsilon_4 z^2)^2 \Pi_{2,2}^{\mathrm{Tr}} }{\mathcal{D}_{k_s}^d}  +(\varepsilon_1 \cdot \varepsilon_2 \varepsilon_3 \cdot \varepsilon_4)^2 \Pi_{2,0} \nonumber \\
    &+2(\varepsilon_1 \cdot \varepsilon_2 \varepsilon_3 \cdot \varepsilon_4 z^2 \Pi_{1,1}+z^2 W_s)(\varepsilon_1 \cdot \varepsilon_2 \varepsilon_3 \cdot \varepsilon_4 \Pi_{1,0} +V_c^s) +((V_c^s)^2+2\varepsilon_1 \cdot \varepsilon_2 \varepsilon_3 \cdot \varepsilon_4 V_c^s \Pi_{1,0})(z^2 k_s^2+4s_{12}s_{34}) \nonumber \\
    &+\mathcal{P}(2,3,4).\label{eq:4pt-graviton}
\end{align}

First of all, we want to emphasize that unlike the usual bulk calculation on spinning particles in AdS which involves complicated bulk integral in axial gauge\cite{Albayrak:2019yve,Albayrak:2018tam}, all of our calculations are just scalar integrals, which can be easily automated by Mathematica. Now by inverse Mellin transform:
\begin{align}
\mathcal{I}_{2,2}&=\frac{z^4\Pi_{2,2}}{\mathcal{D}_{k_s}^{d}} \to \Pi_{2,2}\int \frac{dz}{z^{d+1}} (z^2 \phi_1 \phi_2)(\mathcal{D}_{k_s}^d)^{-1}(z^2 \phi_3 \phi_4), \\
\mathcal{I}_{2,1}&=\Pi_{2,1} \to \frac{\Pi_{1,1}}{k_s^2}\int \frac{dz}{z^{d+1}}z^2(\partial_{z1}-\partial_{z2})(\partial_{z3}-\partial_{z4})\phi_1 \phi_2 \phi_3 \phi_4.
\end{align}
where $\partial_{zi}$ means the $\partial_z$ acting on the corresponding leg only. The inversion is defined via the standard Green function:
\begin{align}
    (\mathcal{D}(z))^{-1} \mathcal{O}(z)&= \int \frac{dy}{y^{d+1}} G(z,y) \mathcal{O}(y), \\
    \mathcal{D}(z)G(z,y)&=z^{d+1} \delta(z-y).
\end{align}
We will evaluate the integral in $d=3$,
\begin{align}
\mathcal{I}_{2,2}=&\Pi_{2,2}\int \frac{dz}{z^{d+1}} (z^2 \phi_1 \phi_2)G(k_s,z,z')(z'^2 \phi_3 \phi_4) \nonumber \\
=&\Pi_{2,2}\left( \frac{2 k_1 k_2 k_3 k_4 \left(E_L E_R+E k_s\right)}{E_L^2 E^3 E_R^2}+\frac{k_1 k_2 \left(E_L k_{34}+E k_s\right)}{E_L^2 E^2 E_R}+\frac{k_3 k_4 \left(E k_s+E_R k_{12}\right)}{E_L E^2 E_R^2}+\frac{E_L E_R-k_s^2}{E_L E E_R} \right), \\
\mathcal{I}_{2,1}=&\frac{\Pi_{1,1}(k_1-k_2)(k_3-k_4)}{k_s^2} \left( \frac{2 k_1 k_3 k_4 k_2}{E^3}+\frac{k_1 k_2k_{34}}{E^2}+\frac{k_3 k_4 k_{12}}{E^2}+\frac{k_{12}k_{34}}{E} \right) .
\end{align}
The integral for $\Pi_{2,0}$ clearly involve more $z$ derivatives, but it is essentially just contact diagram, we will not present the integrated expression here, but we have explicitly verified that agree with \cite{Bonifacio:2022vwa}. In particular, we matched our $\Pi_{2,0}$ with $f_{(2,0)}^{(s)}(E_LE_R-sk_T)\Pi_{2,0}^{(s)}+f_c$ in Eq(2.39) \cite{Bonifacio:2022vwa}. \\
Moving forward, we can utilize the formula in \cite{Bonifacio:2022vwa,Baumann:2020dch}, which establishes a connection between the wavefunction coefficient and In-In correlator, this will give us the graviton trispecturm.



\end{widetext}

\end{document}